\PassOptionsToPackage{hyphens}{url} 
\documentclass[%
 aip,
 amsmath,amssymb,
 reprint,%
]{revtex4-2}
\usepackage{dcolumn,mathptmx,setspace}
\usepackage{bbm}
\usepackage[colorinlistoftodos]{todonotes}
\usepackage[utf8]{inputenc} 
\usepackage[T1]{fontenc}    

\DeclareMathOperator{\E}{\mathbb{E}}

\usepackage[utf8]{inputenc} 
\usepackage[noend]{algpseudocode}

\usepackage{amssymb,amsmath,mathrsfs,stmaryrd,amsthm,mathtools,amsfonts,graphicx,hyperref,url,booktabs,nicefrac,comment,microtype,tikz-cd,lingmacros,nccmath,tree-dvips, bbm, bm, etoolbox,algorithm,caption,subcaption,color,enumitem,relsize,soul}

\usepackage{amsmath}

\begin{document}

\preprint{AIP/123-QED} 
\title[Dual-domain analysis of gun violence incidents in the United States]{Dual-domain analysis of gun violence incidents in the United States}


\author{Nick James}
\affiliation{ 
School of Mathematics and Statistics, University of Melbourne, Victoria, 3010, Australia}%
\affiliation{Melbourne Centre for Data Science, University of Melbourne, Victoria, 3010, Australia}
\author{Max Menzies}
\email{max.menzies@alumni.harvard.edu}
\affiliation{%
Beijing Institute of Mathematical Sciences and Applications, Tsinghua University, Beijing, 101408, China}%

\date{August 15 2022}
\begin{abstract}

This paper applies new and recently introduced approaches to study trends in gun violence in the United States. We use techniques in both the time and frequency domain to provide a more complete understanding of gun violence dynamics. We analyze gun violence incidents on a state-by-state basis as recorded by the Gun Violence Archive. We have numerous specific phenomena of focus, including periodicity of incidents, locations in time where behavioral changes occur, and shifts in gun violence patterns since April 2020. First, we implement a recently introduced method of spectral density estimation for nonstationary time series to investigate periodicity on a state-by-state basis, including revealing where periodic behaviors change with time. We can also classify different patterns of behavioral changes among the states. We then aim to understand the most significant shifts in gun violence since numerous key events in 2020, including the COVID-19 pandemic, lockdowns, and periods of civil unrest. Our dual-domain analysis provides a more thorough understanding and challenges numerous widely held conceptions regarding the prevalence of gun violence incidents.

\end{abstract}

\maketitle

\begin{quotation}

Especially in the leadup to the 2022 election, gun violence remains a high-profile and challenging issue in the United States (U.S.). The U.S. leads the world both in gun ownership and prevalence (the only country with more guns than people \cite{small_arms_survey}) and leads the developed world in gun incidents, injuries, homicides and suicides.\cite{hemenway_homicide,IHME_healthdata} The challenge of gun violence is a highly charged political issue both on the federal and state levels.\cite{spitzer_politics_2020} Gun regulation varies significantly between states,\cite{stategunlaws} with a substantial political discourse around the utility or performance of increased regulation on a state level to decrease the incidence of violence.\cite{rodriguez_andres_gun_2011} Recent events such as the COVID-19 pandemic, resulting lockdowns, and an increased interest in social justice and police reform should call for increased research into trends in gun incidents, particularly on a state-by-state basis. This paper meets this need through a thorough and mathematically original investigation into numerous distinct features of the U.S. gun violence epidemic, with a focus on state-by-state differences, collective trends, and changes since 2020.

\end{quotation}

\section{Introduction}

The Second Amendment of the U.S. Constitution grants Americans a unique level of firearm access among the developed world.\cite{2ndamendmentunique} The right to bear arms remains a longstanding and important legal and cultural principle to the American people.\cite{gunrights} Anchored in this principle, gun ownership is much less regulated and prevalent in the U.S. than in similarly developed countries around the world. Political opinions about the nature of gun ownership vary sharply, with numerous citizens and states differing in their political opinions and enacted regulations around the sale, carry and ownership of guns.\cite{galluppolls} While legal gun use remains a significant part of U.S. culture, particularly in more rural locations,\cite{pew_rural} illegal and violent use of firearms creates numerous tragic occurrences, some of high profile, and others barely reported.\cite{mazeika_effect_2022}

Due to U.S. regulations around studying gun violence as a public health issue,\cite{atlantic_gunshealth} there is comparatively less analysis on trends in gun incidents than one might expect. In particular, there is a dearth of in-depth mathematical and statistical analyses of gun incidents on a state-by-state basis, particularly that incorporates recent data and the overall surge of gun violence since 2020. Related existing mathematical research exists on patterns in the acquisition of guns across the U.S.\cite{wang_spatiotemporal_2022} Given the significance of legal gun ownership to U.S. culture and the political significance of the ongoing debates around gun regulation,\cite{npr_polling} increased mathematical analysis of gun violence remains necessary to ground the debate and inform the public. In addition, mathematical analysis of trends in gun violence could provide numerous benefits to policymakers, allowing them to track the most successful policies regarding gun violence reduction, both between states and over time.

For this end, we aim to conduct an original and thorough analysis on gun violence incidents on a state-by-state basis through the perspective of \emph{multivariate time series analysis}. We draw upon numerous approaches that have been successfully applied in numerous disparate fields such as epidemiology,\cite{james2021_CovidIndia,Li2021_Matjaz,Manchein2020,Blasius2020,james2021_TVO,Perc2020,jamescovideu}  finance,\cite{james2022_stagflation,Drod2021_entropy,james_georg,Wtorek2021_entropy,james_arjun} cryptocurrency\cite{Sigaki2019,Drod2020_entropy,James2021_crypto2,Drod2020,Wtorek2020} and other fields.\cite{Ribeiro2012,james2021_hydrogen,Merritt2014,james2021_olympics,Clauset2015} In particular, we draw inspiration from the fields of Bayesian nonstationary time series analysis,\cite{james2021_spectral,Dahlhaus1997,james2021_MJW} distance analysis, \cite{Moeckel1997,James2021_geodesicWasserstein,Mendes2019} and clustering\cite{Machado2020,James2021_virulence} which have been widely applied. This is part of a broader push in the literature to apply techniques from mathematical and statistical physics to study societal phenomena, termed social physics.\cite{Perc_social_physics} Within social physics, researchers have analyzed crime and terrorism \cite{Helbing2014,Perc2013} but not specifically U.S. gun violence, as far as we are aware.

This paper aims to study mathematical properties of trends in gun violence incidents on a state-by-state basis, drawing upon the aforementioned frameworks. Section \ref{sec:methodology} describes the methodology used in this paper, primarily spectral density estimation of locally stationary time series in a Bayesian framework. In Section \ref{sec:Adaptspec}, we study the nonstationary structure of state time series with a focus on time-varying spectral density of gun events. We use a robust framework of two distinct distances to reveal similarity between U.S. states in their periodic structure as well as outlier states. In Section \ref{sec:prepost}, we focus more specifically on changes in behavior since April 2020, both on a temporal and spectral basis. We conclude in Section \ref{sec:conclusion}.

\section{Methodology and background on spectral density estimation}
\label{sec:methodology}

In this section, we provide a terse overview of spectral density estimation of stationary and nonstationary time series and describe the specific methodology we use to analyze the data in this paper.

The basic building block of the theory is the \emph{stationary time series} $X_t,t=1,...,n.$ By definition, this is a sequence of integrable random variables $X_t$ in which the autocovariance function $\gamma(k)=\E[(X_{t}-\mu)(X_{t+k}-\mu)]$ only depends on $k$. Spectral density estimation concerns the study of the associated power spectral density function (PSD), defined as follows: 
 \begin{equation}
     f(\nu) = \sum^{\infty}_{k=-\infty} \gamma (k) \exp (-2\pi i\nu k), \text{ for} -\frac{1}{2} \leq \nu \leq \frac{1}{2}.
 \end{equation}
In particular, the \emph{Fourier frequencies} are defined as $\nu_j=\frac{j}{n}, j=0,1,...,n-1$. By symmetry, we can restrict attention to $j=0,1,...,m-1$ where $m=[\frac{n}{2}]+1$.  The \emph{discrete Fourier transform} (DFT) of the time series is defined as follows:
\begin{equation}
Z(\nu_j) =\frac{1}{\sqrt{n}} \sum_{t=1}^n X_{t} \exp(-2 \pi i \nu_j t), \text{ for } j = 0, ... ,n-1.
\end{equation}
For each Fourier frequency, a noisy but unbiased estimate of the PSD $f(\nu_{k})$ is the \emph{periodogram}, \cite{Choudhuri2004} defined by $I(\nu_{j})=|Z(\nu_j)|^2$. Based on theoretical properties\cite{Brockwell1991}, previous researchers established a signal plus noise representation of the periodogram:\cite{Rosen2009,Rosen2012}
\begin{equation}
\label{eq:log_periodigram}
\log I(\nu_{j}) = \log f(\nu_{j}) + \epsilon_{j}, 
\end{equation}
where $\epsilon_{j}$ are (the log of) exponentially distributed noise errors. The Whittle likelihood function \cite{Whittle1957,Whittle1954} provides an estimation of the likelihood of the log periodogram given the true power spectrum. It is defined as follows:
\begin{equation}
\label{eq:stationarywhittle}
p(\mathbf{I}| \mathbf{f}) = (2 \pi)^{-m/2} \prod_{j=0}^{m-1} \exp\left({-\frac{1}{2}\left[\log f(\nu_{j}) + \frac{I(\nu_{j})}{f(\nu_{j})}\right]}\right).
\end{equation}

To generalize the above theory and framework to nonstationary processes, we assume that every time series analyzed in this paper can be described as a \emph{Dahlhaus piecewise stationary process} \cite{Dahlhaus1997}. Specifically, let a time series of length $T$, $(Y_t)_{t=1,...,T}$, consist of an unknown number of segments $r$ and change points $\tau_1,...,\tau_{r-1}$ between segments. For notational convenience, set $\tau_0=0, \tau_r=T$. Then, the entire time series $Y_t$ can be written as
\begin{align}
    \label{eq:nonstationarymodel}
    Y_t = \sum_{i=1}^r \bm{1}_{[\tau_{i-1}+1,\tau_i]} X_t^i,
\end{align}
where each $X_t^i$ is an independent and stationary time series over the interval $[\tau_{i-1}+1,\tau_i]$. Following the notation for the stationary setting, let $n_i = \tau_i - \tau_{i-1}$ be the length of the $i$th segment, $m_i  =[\frac{n_i}{2}]+1$ be the effective length of each periodogram, and let $\mathbf{f}_i \in \mathbb{R}^{m_i}$ denote the PSD of the stationary time series $X_t^i$. By independence of the processes, the Whittle likelihood approximation of the nonstationary model for a given partition $\boldsymbol{\tau}=(\tau_1,...,\tau_{r-1})$ is as follows:
\begin{align}
    \label{eq:nonstationarywhittle}
    L(\mathbf{f}_1,...,\mathbf{f}_r | \boldsymbol{\tau}, \mathbf{I}_1,...,\mathbf{I}_m) = \prod_{i=1}^r p(\mathbf{I}_i| \mathbf{f}_i),
\end{align}
where each local Whittle likelihood $p(\mathbf{I}_i| \mathbf{f}_i)$ is calculated according to (\ref{eq:stationarywhittle}).

In our paper, we analyze nonstationary time series $y_t$ of gun violence events in a particular state. The heart of our methodology is a \emph{reversible jump Markov chain Monte Carlo} algorithm to estimate partition points $\tau_i$ and estimate the PSD $\mathbf{f}_i$ of the stationary segments $[\tau_{i-1}+1,\tau_i]$. This scheme partitions the time series and models each local log PSD $\log f(\mathbf{\nu})$ with a Gaussian process \cite{Rasmussen2006} with covariance matrix $\Omega$. A flexible and optimized eigendecomposition is then utilized to produce an estimate for the log PSD, iterated within the sampling scheme. This method is referred to as an \emph{optimized smoothing spline}: further details on the methodology and sampling scheme can be found in Ref. \onlinecite{james2021_spectral}.

Perhaps even more important than estimating the local power spectral density functions, this whole methodology also determines the location of the change points $\tau_i$, recording uncertainty around their location in the form of a probability distribution. Specifically, after discarding burn-in, the sampling scheme produces a distribution over the number of change points $r$, and conditional on $r$, a set of $r$ points with uncertainty intervals. We select the maximally likely number of change points $r_0$ for each individual time series, resulting in a set of $r_0$ distributions that describe where the change points may fall. 

Following Ref. \onlinecite{james2021_MJW}, we term this second output a \emph{set with uncertainty}, defined as a finite set of probability distributions $\tilde{S}=\{f_1,...,f_{r_0}\}$ with disjoint supports. In this formulation, each $f_i$ tracks the distribution of the $i$th change point $\tau_i$, recording its uncertainty. As a remark, when each $f_i$ is simply a Dirac delta distribution $\delta_{x_i}$ (that is, there is no uncertainty around the change points), then the set with uncertainty $\tilde{S}$ reduces to just a
normal set of points $S=\{x_1,...,x_{r_0}\}$. The procedure always returns a change point at $\tau=T$, so every $\tilde{S}$ is non-empty.


In this paper, we will seek to quantify discrepancy between different states' sets of change points (with uncertainty considered). As such, we use the \emph{MJ-Wasserstein semi-metric}\cite{james2021_MJW} between sets with uncertainty. This is defined as follows:
\begin{align}
\label{eq:MJW}
    d^p_{MJW}(\tilde{S},\tilde{T}) &= \Bigg(\frac{\sum_{g\in \tilde{T}} d_W(g,\tilde{S})^p}{2|\tilde{T}|} + \frac{\sum_{{f} \in \tilde{S}} d_W(f,\tilde{T})^p}{2|\tilde{S}|} \Bigg)^{\frac{1}{p}}, 
\end{align}
where $\tilde{S},\tilde{T}$ are (non-empty) sets with uncertainty and $p \geq 0$ is a parameter. In this paper, we set $p=1$ throughout.

This allows us to produce a distance measure between different states' gun violence event time series. Simply, if $(Y^{(i)}_t), i=1,...,N$ is a collection of $N$ time series over a period $t=1,...,T$, then we can apply our methodology to produce sets with uncertainty $\tilde{S}_i$ and then analyze discrepancy in their change points (incorporating uncertainty) via the $N \times N$ distance matrix $D_{ij}=d^1_{MJW}(\tilde{S}_i,\tilde{S}_j)$.

\section{Results of spectral density estimation and change point analysis}
\label{sec:Adaptspec}

The data studied in this paper are drawn from the Gun Violence Archive.\cite{gundata} We simply analyze the daily number of recorded events $Y_t^{({i})}$ (incidents of illegal shootings) on a state-by-state basis, $i=1,...,N,t=1,...,T$, in each of the $N=51$ U.S. states and D.C. Our data ranges January 1, 2018 to June 9, 2022, a period of $T=1621$ days. All the methodologies of Section \ref{sec:methodology} is applied to these time series.

\begin{figure*}
    \centering
    \begin{subfigure}[b]{0.49\textwidth}
        \includegraphics[width=\textwidth]{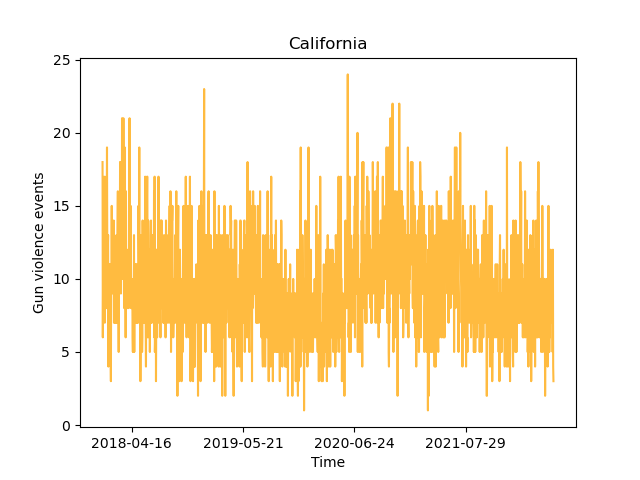}
        \caption{}
        \label{fig:California}
    \end{subfigure}
    \begin{subfigure}[b]{0.49\textwidth}
        \includegraphics[width=\textwidth]{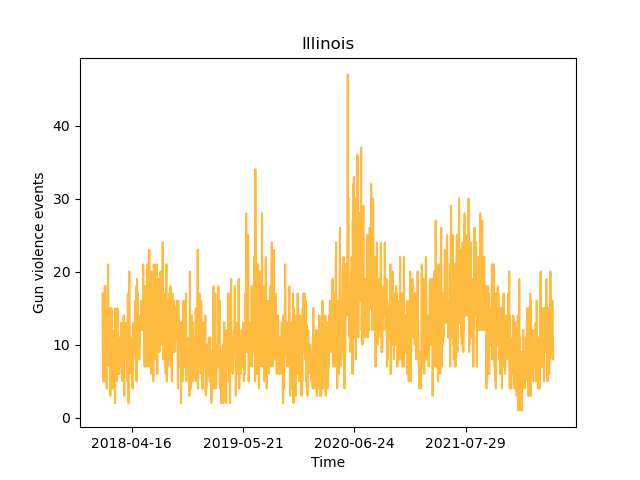}
        \caption{}
        \label{fig:Illinois}
    \end{subfigure}
\begin{subfigure}[b]{0.49\textwidth}
        \includegraphics[width=\textwidth]{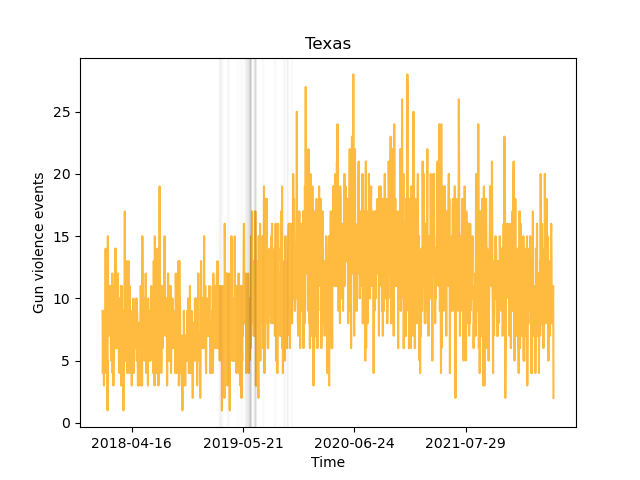}
        \caption{}
        \label{fig:Texas}
    \end{subfigure}
\begin{subfigure}[b]{0.49\textwidth}
        \includegraphics[width=\textwidth]{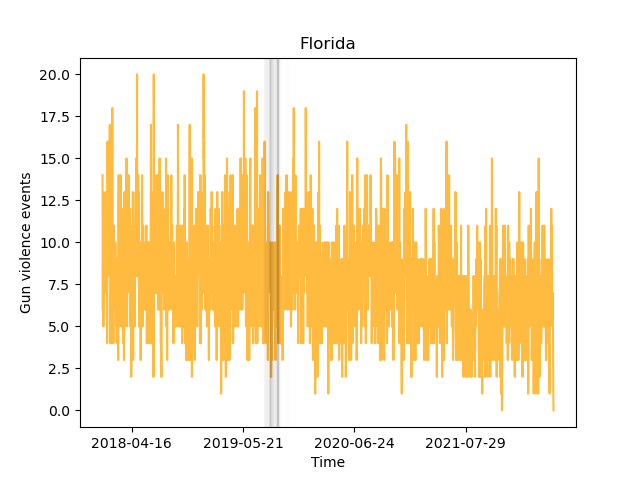}
        \caption{}
        \label{fig:Florida}
    \end{subfigure}
\begin{subfigure}[b]{0.49\textwidth}
        \includegraphics[width=\textwidth]{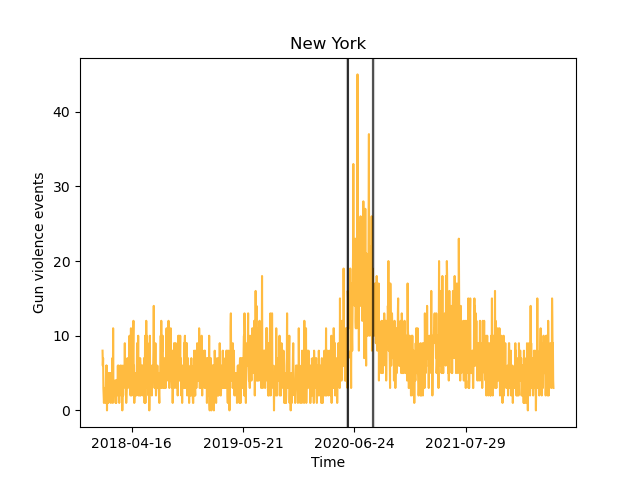}
        \caption{}
        \label{fig:NewYork}
    \end{subfigure}
\begin{subfigure}[b]{0.49\textwidth}
        \includegraphics[width=\textwidth]{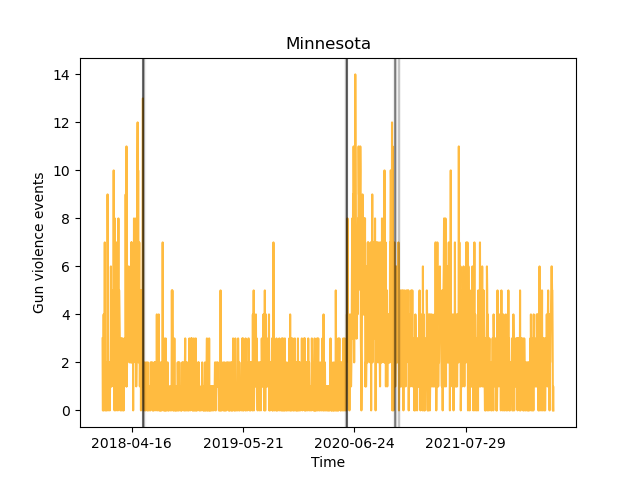}
        \caption{}
        \label{fig:Minnesota}
    \end{subfigure}
    \caption{Time series with annotated change points with uncertainty. Figures (a) and (b) display California and Illinois, neither of which contain a change point during our analysis window. Figures (c) and (d) show the time series of Texas and Florida, respectively. Both figures show that these states contain only one change point; however, these states exhibit very different shifts in their respective behaviors. Figures (e) and (f) show New York and Minnesota, where there is a pronounced spike in gun violence activity in mid-2020. Collectively, these figures show substantial heterogeneity in the behavioral changes in U.S. gun violence on a state-by-state basis, and that shifts during 2020 were far from uniformly felt across the nation.}
   \label{fig:temporal_time_series}
\end{figure*}

We begin by visualizing several states' gun violence event time series with annotated change points as determined by our methodology. First, California (Figure \ref{fig:California}) and Illinois (\ref{fig:Illinois}) exhibit obvious periodic waves - but display no obvious change point in gun violence behaviors. Next, Texas (\ref{fig:Texas}) and Florida (\ref{fig:Florida}) both contain one change point, determined to be in the latter part of 2019. However, they have markedly different explanations for their respective change points, with Florida experiencing a decrease in mean gun violence events, while Texas shows a significant increase (and a slight subsequent decrease) in both the mean and variance of incidents. Third, New York (\ref{fig:NewYork}), Michigan, Missouri, Ohio, and other states exhibit two change points each, corresponding to a sharp spike gun violence activity throughout mid-2020. This timing is consistent with a broad rise in U.S. crime and unrest in the wake of COVID-19 lockdowns, a spike in unemployment, the murder of George Floyd, and associated protests.\cite{crimespikeUS,homicideratesUS,meyer_changes_2022,premkumar_intensified_2020}

\begin{figure*}
    \centering
    \includegraphics[width=\textwidth]{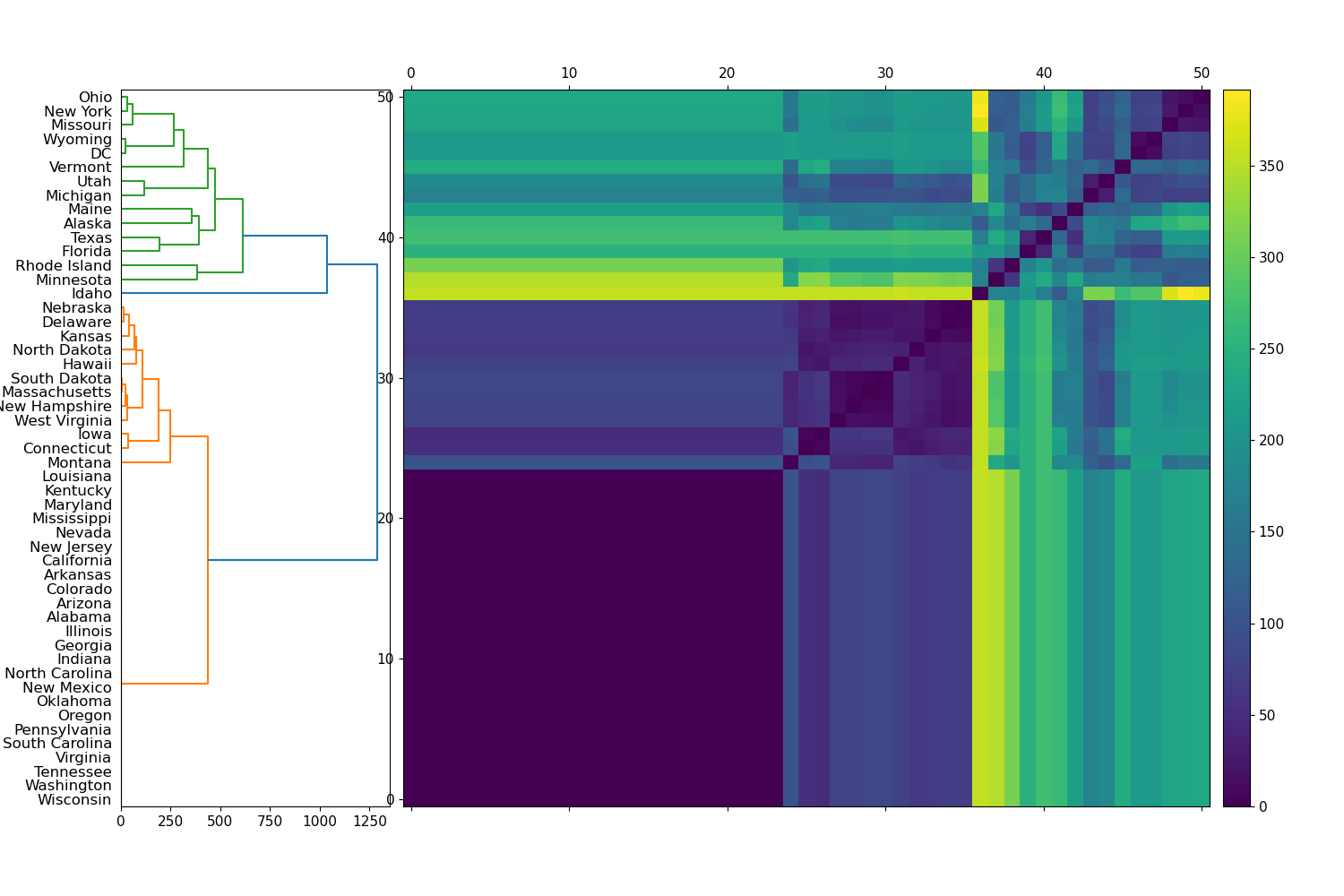}
    \caption{Hierarchical clustering applied to the distance matrix $D^{MJW}$. There are three clear clusters. The bottom cluster contains states that do not possess a change point, indicating no obvious change in the periodic nature of gun violence behaviors throughout our analysis window. Cluster two above that consists mostly of states with a surprising change point late in our analysis window. The top cluster has the lowest similarity, but broadly, these states have at least one early change point. The states at the very top of the third cluster are characterized by a ``2020 summer jump'' in violence. Broadly speaking, one again sees substantial heterogeneity between states with respect to their shifts in gun violence behavior, but substantial structure in this heterogeneity can be gleaned. }
    \label{fig:MJWdend}
\end{figure*}

Having identified the number and precise location of changes in gun violence behavior for each state, we use the distance (\ref{eq:MJW}) and apply hierarchical clustering to the resulting distance matrix. Figure \ref{fig:MJWdend} displays the similarity of each state's change point profile, and reveals three primary clusters. The most trivial cluster (the bottom of Figure \ref{fig:MJWdend}) consists of states with no detected change points throughout the entire window of analysis, including California and Illinois. The second cluster (above the bottom one) spans states from Nebraska to Montana. The most defining attribute of these states is the existence of a (quite unexpected) change point relatively late in the analysis window. For the most part, this structural break corresponds to August-September 2021 and coincides with a clear reduction in the average daily number of gun violence events.

The final cluster (on top) displays the least collective similarity, with relative heterogeneity exhibited between the state's change point profiles. This is confirmed by the dendrogram's tree, where neighboring states are clearly shown to be less similar with the more diffuse (green) leaf. However, some clear and interesting patterns can be discerned. Three states, Ohio, New York, and Missouri, display the clearest ``2020 summer jump'', with two change points marking an increase and decrease in incidents of gun violence at the start and end of the 2020 summer, respectively, represented by New York in Figure \ref{fig:NewYork}. Utah and Michigan feature the same increase in early 2020, and a second change point (marking a subsequent decrease) that occurs in late 2021; Minnesota (\ref{fig:Minnesota}) features an earlier change point in mid-2018 and then a 2020 summer jump. Other states in this more disparate cluster, such as Texas and Florida, generally exhibit at least one early change point (notably Idaho, whose single change point is earlier than any other state). On the other hand, the two aforementioned primary clusters are defined by late change points or none at all.

\begin{figure*}
    \centering
    \begin{subfigure}[b]{0.49\textwidth}
        \includegraphics[width=\textwidth]{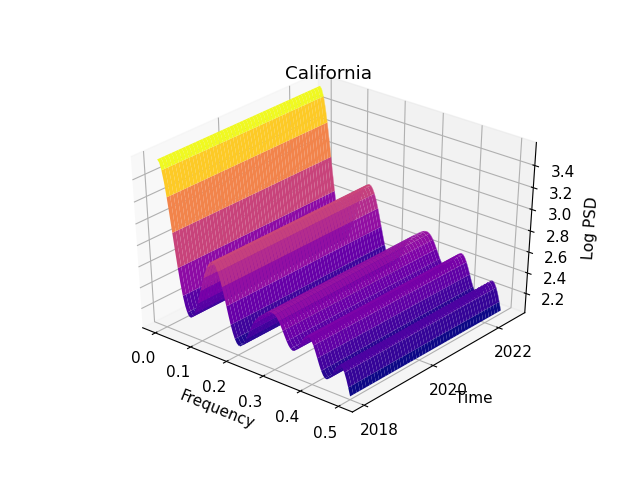}
        \caption{}
        \label{fig:California_adaptspec}
    \end{subfigure}
    \begin{subfigure}[b]{0.49\textwidth}
        \includegraphics[width=\textwidth]{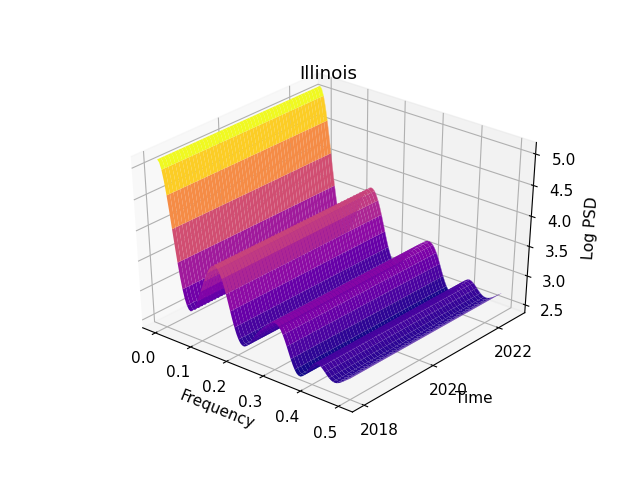}
        \caption{}
        \label{fig:Illinois_adaptspec}
    \end{subfigure}
\begin{subfigure}[b]{0.49\textwidth}
        \includegraphics[width=\textwidth]{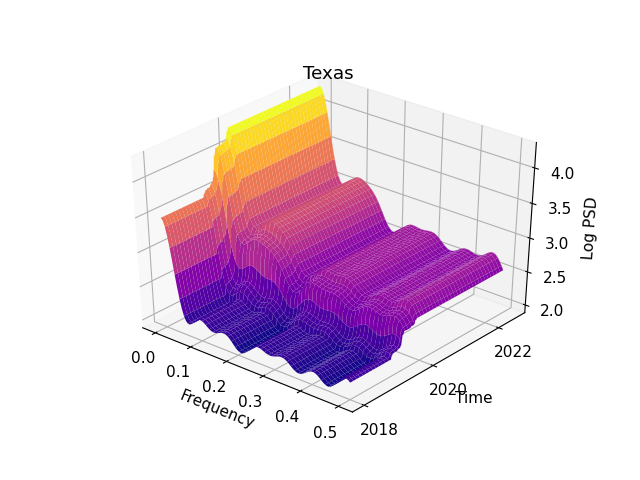}
        \caption{}
        \label{fig:Texas_adaptspec}
    \end{subfigure}
\begin{subfigure}[b]{0.49\textwidth}
        \includegraphics[width=\textwidth]{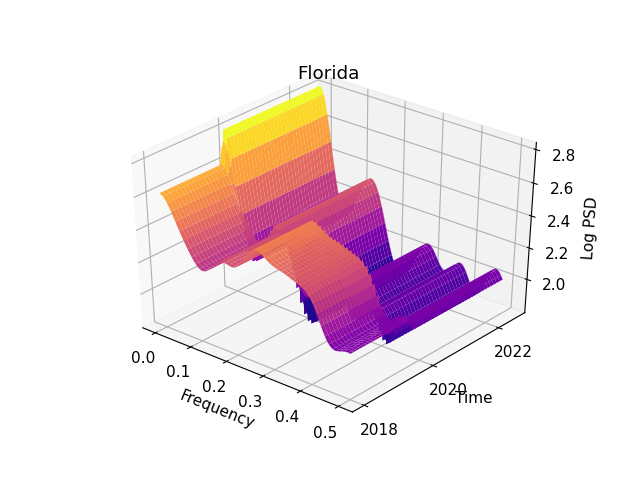}
        \caption{}
        \label{fig:Florida_adaptspec}
    \end{subfigure}
\begin{subfigure}[b]{0.49\textwidth}
        \includegraphics[width=\textwidth]{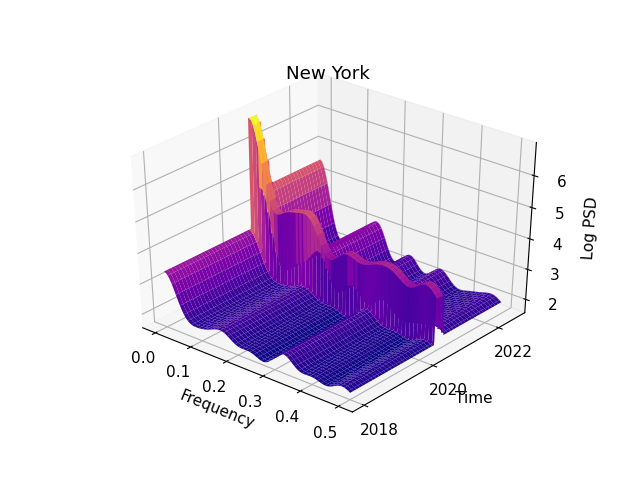}
        \caption{}
        \label{fig:New_york_adaptspec}
    \end{subfigure}
\begin{subfigure}[b]{0.49\textwidth}
        \includegraphics[width=\textwidth]{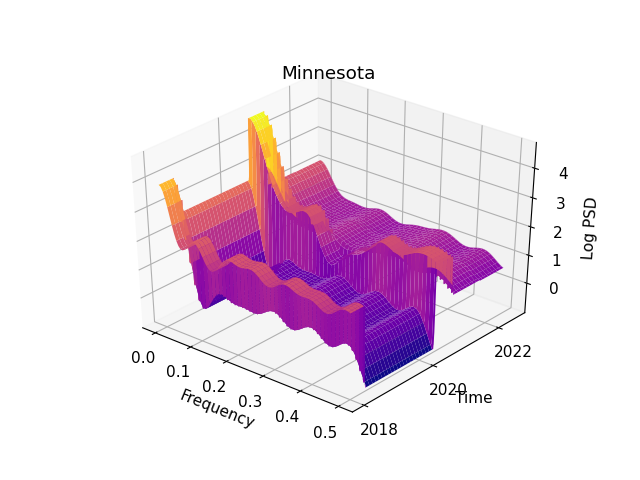}
        \caption{}
        \label{fig:Minnesota_adaptspec}
    \end{subfigure}
    \caption{Time-varying adaptive spectral densities corresponding to (a) California, (b) Illinois, (c) Texas, (d) Florida, (e) New York and (f) Minnesota. California and Illinois display no structural break and a consistent power spectral density throughout the period analyzed. Texas and Florida both exhibit changes in the structure of the power spectrum; however, it is clear that there is a more obvious change in Florida's prominent periodic components. Finally, the surfaces of New York and Minnesota both display a clear spike in amplitude and spectral complexity during mid-2020. Again, these figures show substantial heterogeneity in the changes in U.S. gun violence frequency behavior on a state-by-state basis, with some states exhibiting notable changes in their estimated power spectral density over time; yet, these shifts were far from uniform.}
   \label{fig:Adaptspecs}
\end{figure*}

In Figure \ref{fig:Adaptspecs}, we display the time-varying estimated power spectral density for the states initially shown in Figure \ref{fig:temporal_time_series}. California and Illinois, shown in Figures \ref{fig:California_adaptspec} and \ref{fig:Illinois_adaptspec}, respectively, clearly show no change in the state's log PSD, which is consistent with the initial finding that neither state has a marked change in gun violence behaviors. Texas and Florida (Figures \ref{fig:Texas_adaptspec} and \ref{fig:Florida_adaptspec}) clearly display shifts in the power spectrum in the latter part of 2019. Texas primarily exhibits a translation upward (consistent with greater total gun violence activity), while Florida displays a markedly different spectrum - indicating a major shift in the periodic nature of the state's gun violence. Finally in Figures \ref{fig:New_york_adaptspec} and \ref{fig:Minnesota_adaptspec}, New York and Minnesota's mid-2020 spike in gun violence activity is clearly observed. During this period, we see a consistent display of gun violence exhibiting high power at frequencies corresponding to quarterly, weekly and biweekly periodic cycles.

\section{Change in gun violence incidence before and after April 2020}
\label{sec:prepost}

In this section, we take a closer and more specific investigation of the changes to each state's pattern of gun violence incidents around April 2020. We perform a direct before vs after comparison regarding both temporal and spectral behaviors. This date reflects a confluence of events that may have significantly influenced U.S. violence, including the declaration of COVID-19 as a global pandemic on March 11, 2020,\cite{UN_pandemic} an immediate market crash,\cite{CNBC} the imposition of lockdowns and business closures in every state,\cite{first_US_lockdowns} an associated spike in unemployment,\cite{unemploymentspike} and shortly afterward the murder of George Floyd and the resulting summer of racial justice protests.\cite{GeorgeFloyd} With this aim in mind, we analyze the difference in each state time series' mean and amplitude-adjusted power spectral density (PSD) function pre- and post-April 2020. The former highlights shifts in the total amount or amplitude of gun violence crime, while the latter identifies time series with meaningfully different periodicities - indicating potential differences in gun violence behaviors on a state-by-state basis.

\begin{figure*}
    \centering
    \begin{subfigure}[b]{0.49\textwidth}
        \includegraphics[width=\textwidth]{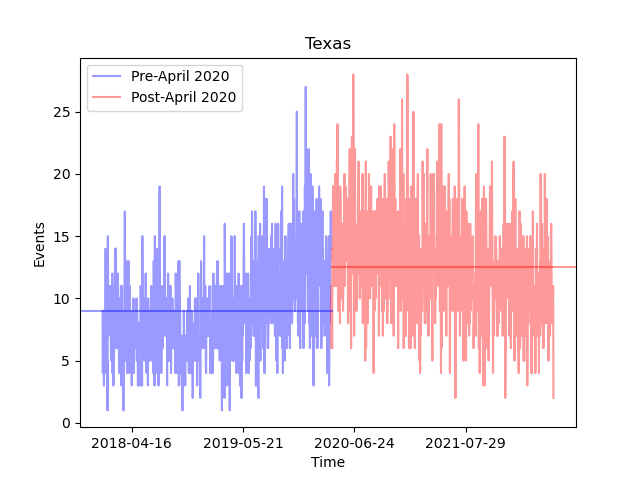}
        \caption{}
        \label{fig:Texas_temporal}
    \end{subfigure}
    \begin{subfigure}[b]{0.49\textwidth}
        \includegraphics[width=\textwidth]{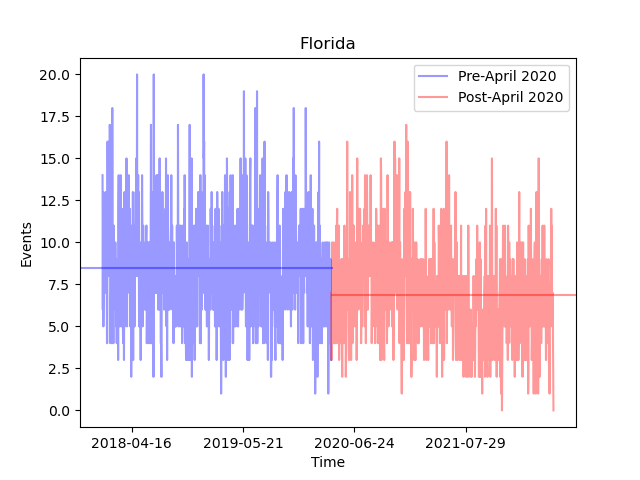}
        \caption{}
        \label{fig:Florida_temporal}
    \end{subfigure}
\begin{subfigure}[b]{0.49\textwidth}
        \includegraphics[width=\textwidth]{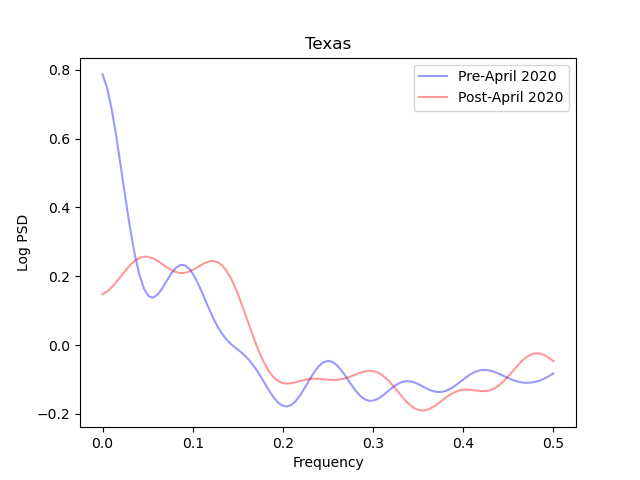}
        \caption{}
        \label{fig:Texas_spectral}
    \end{subfigure}
\begin{subfigure}[b]{0.49\textwidth}
        \includegraphics[width=\textwidth]{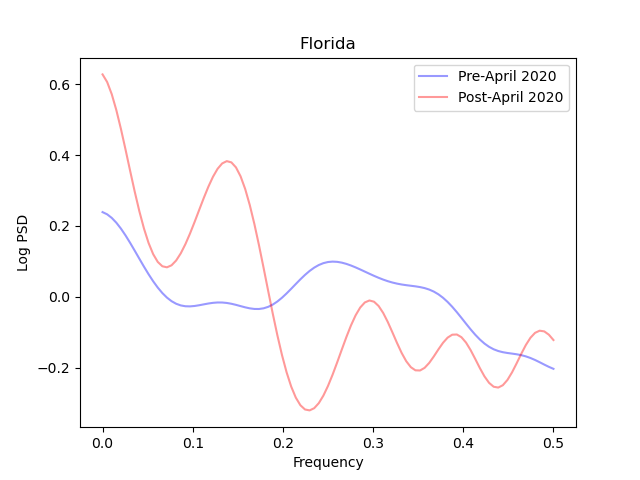}
        \caption{}
        \label{fig:Florida_spectral}
    \end{subfigure}
\begin{subfigure}[b]{0.49\textwidth}
        \includegraphics[width=\textwidth]{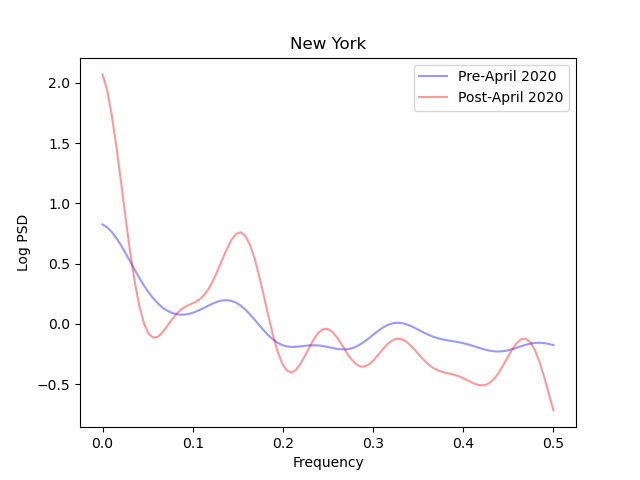}
        \caption{}
        \label{fig:NewYork_spectral}
    \end{subfigure}
\begin{subfigure}[b]{0.49\textwidth}
        \includegraphics[width=\textwidth]{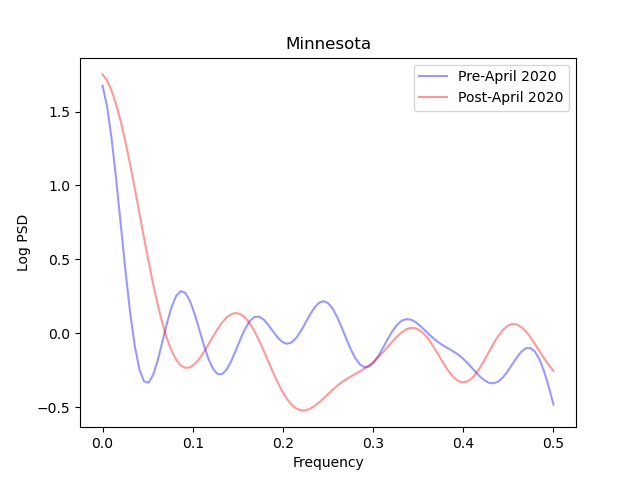}
        \caption{}
        \label{fig:Minnesota_spectral}
    \end{subfigure}
    \caption{(a) and (b) show Texas and Florida's time series, emphasizing the change in amplitude of gun violence before and after April 1, 2020. There is a clear shift upward in Texas' mean daily gun violence activity, and a clear shift downward for Florida. (c) and (d) show the power spectral density for each period. There is a sharp reduction in Texas' low frequency components in the period proceeding April 2020, while Florida displays much greater power at a frequency $\nu \sim 0.14$. This would suggest that in the earlier period, Texas exhibits less gun violence on an annual, biannual and quarterly periodicity, while Florida produces more gun violence with a weekly periodicity. (e) and (f) show the analogous plots for New York and Minnesota. New York exhibits much higher power at low frequency components, while Minnesota displays clear deterioration at a weekly component after April 2020. Collectively, these figures highlight the benefit of our dual-domain analysis: shifts in behavior pre/post-2020 may be quite different in the temporal and spectral domain.}
   \label{fig:Temporal_spectral_deviation}
\end{figure*}

In Table \ref{tab:temporal}, we compute and display $\mu_{post} - \mu_{pre}$ for each state, the difference in mean daily gun violence events before and after April 1, 2020. There is a relatively even distribution of states moving upward and downward in their gun violence activity, with 29 of the states analyzed experiencing a decrease in their mean gun violence events and 22 (including Washingon, D.C.) experiencing an increase. The most notably positive changes belong to Texas, New York, Illinois, Michigan and Minnesota, displaying an average increase in average gun violence events per day of 3.49, 3.45, 3.12, 2.13 and 1.36, respectively. Although these states share a meaningful increase in gun violence activity, there is marked variability in the general trajectory beyond April 2020. The most idiosyncratic trajectory is Texas (Figure \ref{fig:Texas_temporal}), which actually exhibits some slow decline in violence after the spike in mid-2020. New York, Illinois, Michigan, and Minnesota all exhibit prominent spikes in gun violence activity during mid-2020, often accompanied by increases in the volatility of gun violence events. The five states with the most significant declines in gun violence activity are Florida, Massachusetts, Oklahoma, Arkansas and New Jersey, with mean reductions of -1.61, -1.55, -0.46, -0.40 and -0.35 events per day. Again, there is a fair degree of variability among the state's trajectories. Florida (\ref{fig:Florida_temporal}) exhibits a steady decline in average gun violence events throughout the entire window of analysis, as do Oklahoma and Arkansas, albeit in a more subdued manner. The most interesting feature in Massachusetts and New Jersey's gun violence trends is that, in addition to the steady decline in gun violence events, there is an abrupt reduction corresponding to the latter part of 2021. Although the difference in means highlights the change in gun violence amplitude across U.S. states, we wish to understand whether there has been a meaningful change in gun violence behaviors surrounding our break point.

We now turn to our second piece of analysis, where we wish to study changes in the periodic nature of each state's gun violence activity. Here, we wish to specifically remove consideration of amplitude in the before/after comparison. For this purpose, we use our methodology (Section \ref{sec:methodology}) to obtain an estimate of the log PSD function $\log f(\nu), 0 \leq \nu \leq \frac12$ for each state, treating the before and after April 2020 periods as different time series. In order to appropriately normalize each log PSD (that is, remove amplitude), we consider a linear transformation of time series $X'_t = aX_t + b$. Consulting Section \ref{sec:methodology}, the autocovariance function transforms as $\gamma_{X'}=a^2 \gamma_X$, and hence, the log PSD transforms as $\log f_{X'}= \log(a^2) + \log f_{X}$. Thus, a rescaling of the absolute values of the counts corresponds to an additive constant in the log PSD function. With this in mind, we can appropriately compare log PSD functions by simply mean-adjusting the log PSD, namely considering $g(\nu)= \log f(\nu) - \mu_{\log f(\nu)}.$ We may then perform an amplitude-adjusted before/after comparison by considering the $L^1$ differences between mean-adjusted log PSD functions, $\Delta_i=\| g_{i,post} - g_{i,pre} \|$ for each state $i$. Furthermore, we may take a more focused examination of the before/after spectral difference by computing a modified difference $\epsilon_i = |\sum_j g_{i,post}(\nu_j) - g_{i,pre}(\nu_j)  |$ where the difference is only computed with respect to \emph{key frequencies} (corresponding to a yearly, monthly, weekly, and other periodicities in the frequency domain).

Our frequency analysis shows that since April 2020, Texas' power spectral density (\ref{fig:Texas_spectral}) exhibits significantly less power as low frequencies - signifying less meaningful gun violence contribution on an annual, bi-annual or quarterly periodicity. By contrast, Florida's power spectrum (\ref{fig:Florida_spectral}) indicates the emergence of a powerful annual, bi-annual and weekly gun violence periodicity. The states displaying the most prominent changes in their periodicities are Minnesota, New York, Illinois, Rhode Island and Louisiana (Table \ref{tab:spectral}). Interestingly, each state tells a different story. Minnesota (Figure \ref{fig:Minnesota_spectral}) consists of two interesting properties: both power spectra (before and after) maintain a high degree of power at low frequency components, while the post-April 2020 spectra display clear deterioration at higher frequency components - particularly $\nu \sim 0.14$, corresponding to a 7-day (weekly) cycle. This suggests that annual and biannual gun violence tendencies, representing events such as New Year's Eve, maintained their significance, while weekly gun violence dynamics diminished. New York's power spectrum (\ref{fig:NewYork_spectral}) after April 2020 exhibits much higher power at very low frequency components (corresponding to annual and biannual periodicities) and weekly frequencies. The post-April 2020 power spectrum clearly displays greater complexity, and significantly more variability throughout the range of frequencies. This is indicative of more nuanced periodic gun violence tendencies after April 2020. Illinois displays a broadly similar shape in the power spectrum both pre- and post-April 2020. This is to be expected, as our change point methodology (Section \ref{sec:methodology}) did not detect any break in Illinois' gun violence throughout the entire period of analysis, as mentioned in Section \ref{sec:Adaptspec}. One shift is a translation upward in the power spectrum at lower frequencies, which indicates that post-April 2020 there were more meaningful periodic gun violence tendencies coming from annual, biannual and quarterly event cycles. Similarly, Rhode Island exhibits a much stronger power spectrum at frequency components corresponding to annual, biannual, quarterly and weekly periodicities post-April 2020, when compared to pre-April 2020 gun violence. Finally, Louisiana has two primary findings: the power spectrum is much higher at lower frequencies pre-April 2020, and there is a major upward change in weekly gun violence behaviors post-April 2020.

\section{Conclusion}
\label{sec:conclusion}

In this paper, we study the evolution of U.S. gun violence over the last several years. Importantly, we analyze each state's gun violence in both the temporal and frequency domains, exploring the total change in gun violence incidence as well as the periodic nature at which this gun violence occurs.

We begin by applying a recently introduced method of spectral density estimation for nonstationary time series to determine the temporal location of changes in each state's periodic behaviors. Our methodology is flexible, and does not make any \textit{a priori} assumptions regarding the number or location of each time series' change points. Our framework is capable of modeling time series that exhibit dependence, such as autoregressive processes. This is an important advantage of a method such as ours, which would render many change point detection methodologies unusable on data such as that presented in this paper. Furthermore, given the Bayesian nature of our change point detection methodology, each state's set of change points is accompanied by a distribution over each point's location, which captures the uncertainty in the change. Thus, one can determine whether the probability associated with a state's change in gun violence behaviors is relatively concentrated or diffuse.

Clustering the distance matrix between each state's set of change points with uncertainty allows us to gain insight into different states' gun violence dynamics. Figure \ref{fig:MJWdend} identifies three characteristic state behaviors over our window analysis. Intriguingly, the clusters are remarkably similar to describe: one simply consists of states with no detected breaks, the second consists of states with a relatively unexpected change point quite late in our analysis window, and the final is a more heterogeneous mix characterized by at least one early change point. In particular, the final has a subcluster characterized by marked increases and decreases in gun incidence at the start and end of the 2020 summer, a characteristic ``2020 summer jump.'' However, despite media reporting that this was the case across the U.S., this is by no means the most common behavior.

Given the confluence of major societal events around March-May 2020 (both due to the pandemic and racial justice protests), we conduct an analysis where we specifically focus on changes before and after April 1, 2020. Importantly, we explore shifts in both the temporal and spectral domain, to capture the change in the amplitude of gun violence events, as well as the periodic nature of gun violence activity. With Tables \ref{tab:temporal} and \ref{tab:spectral}, we identify the states with the greatest shifts (in either direction) in amplitude and the greatest change in their periodic profile. This dual-domain analysis provides a more complete picture of changes in gun violence behavior.

From our dual-domain and state-by-state analysis, we have uncovered numerous observations that challenge widely held reporting on U.S. gun violence in recent years. First, it is notable that Minnesota, where George Floyd was killed, and arguably the epicenter of the 2020 racial justice protests,\cite{Minneapolis_protest} exhibited the signal greatest shift in its amplitude-adjusted periodic behaviors after April 2020. Next, we have identified numerous states with a ``2020 summer jump'' in gun violence, as widely discussed by the national media, \cite{crimespikeUS,homicideratesUS} but have shown that this was far from uniform across states. While Minnesota, New York, Ohio and Michigan exhibited such summer jumps, Texas exhibited a rise in gun incidence that took place well before 2020, while Florida has exhibited a welcome and continual decline, also initiated before 2020.

Overall, this paper applies new and recently introduced time and frequency-based time series analysis methodologies to understand the changing dynamics of U.S. gun violence. We believe that this is the first paper in the nonlinear dynamics community to study U.S. gun violence in both the time and frequency domain in a complementary fashion since the COVID-19 pandemic. This work may complement other papers in the literature, such as those studying the periodicity of crime more broadly.\cite{Perc2013} We hope that this work may encourage other researchers to study this topic, and provide insights that may help policymakers to more effectively understand and respond to gun violence across the United States.

\begin{table}
\centering
\begin{tabular}{lr}
  \hline
State & $\mu_{post} - \mu_{pre}$ \\ 
  \hline
Texas & 3.49 \\ 
  New York & 3.45 \\ 
  Illinois & 3.12 \\ 
  Michigan & 2.13 \\ 
  Minnesota & 1.36 \\ 
  Wisconsin & 1.28 \\ 
  Pennsylvania & 1.03 \\ 
  D.C. & 0.66 \\ 
  California & 0.58 \\ 
  Louisiana & 0.57 \\ 
  Arizona & 0.55 \\ 
  Georgia & 0.53 \\ 
  Virginia & 0.51 \\ 
  Colorado & 0.38 \\ 
  Kentucky & 0.32 \\ 
  New Mexico & 0.27 \\ 
  Nevada & 0.18 \\ 
  Indiana & 0.15 \\ 
  Oregon & 0.12 \\ 
  Utah & 0.12 \\ 
  Tennessee & 0.07 \\ 
  Washington & 0.01 \\ 
  Montana & -0.03 \\ 
  South Dakota & -0.03 \\ 
  Hawaii & -0.03 \\ 
  North Carolina & -0.04 \\ 
  Wyoming & -0.04 \\ 
  Idaho & -0.05 \\ 
  North Dakota & -0.06 \\ 
  Ohio & -0.08 \\ 
  Maryland & -0.09 \\ 
  Rhode Island & -0.10 \\ 
  Vermont & -0.10 \\ 
  Missouri & -0.13 \\ 
  Connecticut & -0.15 \\ 
  Nebraska & -0.15 \\ 
  Iowa & -0.17 \\ 
  West Virginia & -0.18 \\ 
  Alabama & -0.18 \\ 
  Delaware & -0.21 \\ 
  New Hampshire & -0.21 \\ 
  Maine & -0.22 \\ 
  Alaska & -0.24 \\ 
  South Carolina & -0.32 \\ 
  Kansas & -0.33 \\ 
  Mississippi & -0.33 \\ 
  New Jersey & -0.35 \\ 
  Arkansas & -0.40 \\ 
  Oklahoma & -0.46 \\ 
  Massachusetts & -1.55 \\ 
  Florida & -1.61 \\ 
   \hline
\end{tabular}
\caption{Difference in mean daily gun violence events before and after April 1, 2020. 22 states (including D.C.) experience an increase, with Texas most of all, and 29 states experience a decrease, most notably  Florida.}
\label{tab:temporal}
\end{table}

\begin{table}
\centering
\begin{tabular}{lrr}
  \hline
State & All frequencies $\Delta_i$ & Key frequencies $\epsilon_i$ \\ 
  \hline
Minnesota & 0.278 & 0.293 \\ 
  New York & 0.240 & 0.608 \\ 
  Illinois & 0.199 & 0.283 \\ 
  Rhode Island & 0.196 & 0.274 \\ 
  Louisiana & 0.194 & 0.204 \\ 
  Florida & 0.185 & 0.272 \\ 
  Pennsylvania & 0.185 & 0.172 \\ 
  Colorado & 0.184 & 0.181 \\ 
  Michigan & 0.181 & 0.182 \\ 
  New Jersey & 0.179 & 0.269 \\ 
  Missouri & 0.169 & 0.329 \\ 
  Oregon & 0.160 & 0.131 \\ 
  Wisconsin & 0.156 & 0.243 \\ 
  California & 0.150 & 0.239 \\ 
  Oklahoma & 0.147 & 0.130 \\ 
  Hawaii & 0.142 & 0.153 \\ 
  Ohio & 0.137 & 0.230 \\ 
  Massachusetts & 0.136 & 0.076 \\ 
  Alabama & 0.130 & 0.086 \\ 
  West Virginia & 0.124 & 0.064 \\ 
  Connecticut & 0.120 & 0.257 \\ 
  Indiana & 0.117 & 0.119 \\ 
  North Carolina & 0.115 & 0.110 \\ 
  Maryland & 0.114 & 0.247 \\ 
  Mississippi & 0.109 & 0.330 \\ 
  Nebraska & 0.104 & 0.076 \\ 
  Vermont & 0.104 & 0.195 \\ 
  D.C. & 0.101 & 0.119 \\ 
  Virginia & 0.100 & 0.162 \\ 
  Georgia & 0.096 & 0.158 \\ 
  Alaska & 0.093 & 0.053 \\ 
  Maine & 0.092 & 0.097 \\ 
  Utah & 0.090 & 0.043 \\ 
  North Dakota & 0.087 & 0.030 \\ 
  Montana & 0.085 & 0.197 \\ 
  Texas & 0.084 & 0.307 \\ 
  New Mexico & 0.084 & 0.071 \\ 
  Tennessee & 0.084 & 0.087 \\ 
  Arizona & 0.079 & 0.070 \\ 
  Delaware & 0.076 & 0.092 \\ 
  South Dakota & 0.072 & 0.181 \\ 
  Nevada & 0.066 & 0.073 \\ 
  Kentucky & 0.065 & 0.132 \\ 
  Iowa & 0.062 & 0.060 \\ 
  Idaho & 0.057 & 0.123 \\ 
  South Carolina & 0.056 & 0.068 \\ 
  Arkansas & 0.054 & 0.159 \\ 
  New Hampshire & 0.053 & 0.068 \\ 
  Washington & 0.051 & 0.069 \\ 
  Wyoming & 0.042 & 0.029 \\ 
  Kansas & 0.024 & 0.020 \\ 
   \hline
\end{tabular}
\caption{$L^1$ norm difference in mean-adjusted log power spectral density functions before and after April 1, 2020, as well as the $L^1$ difference for only key frequencies. Minnesota experiences the greatest shift in its periodic components, indicating substantial changes in its gun violence behavior after events surrounding the death of George Floyd.}
\label{tab:spectral}
\end{table}

\clearpage

\section*{Data availability}
The data that support the findings of this study are openly available at Ref. \onlinecite{gundata}.

\begin{acknowledgments}
The authors would like to thank Jarrah Lacko and James Chok, who helped the authors acquire the data.

\end{acknowledgments}

\bibliography{__newreferences}
\end{document}